\documentclass[12pt]{article}

\usepackage{amssymb}
\usepackage{amsmath}
\usepackage{graphics,graphicx}

\newcommand{\singlespace}{\baselineskip=12pt \lineskiplimit=0pt \lineskip=0pt}
\newcommand{\beq}{\begin{equation}}
\newcommand{\eeq}{\end{equation}}
\newcommand{\lb}{\label}
\newcommand{\beqar}{\begin{eqnarray}}
\newcommand{\eeqar}{\end{eqnarray}}
\newcommand{\bit}{\begin{itemize}}
\newcommand{\eit}{\end{itemize}}
\newcommand{\barr}{\begin{array}}
\newcommand{\earr}{\end{array}}
\def\ds{\displaystyle}

\def\scalp{\mbox{\boldmath $\, \cdot \,$}}
\def\bob{{\, \underline{\overline{\otimes}} \,}}
\def\sop{{\, \overline{\otimes} \,}}
\def\sot{{\, \underline{\otimes} \,}}
\newcommand{\deriv}[2]{\frac{\partial #1}{\partial #2}}

\def\b0{\mbox{\boldmath $0$}}

\def\bA{\mbox{\boldmath $A$}}
\def\bB{\mbox{\boldmath $B$}}
\def\bC{\mbox{\boldmath $C$}}

\def\bE{\mbox{\boldmath $E$}}
\def\bF{\mbox{\boldmath $F$}}

\def\bK{\mbox{\boldmath $K$}}

\def\bP{\mbox{\boldmath $P$}}
\def\bQ{\mbox{\boldmath $Q$}}
\def\bR{\mbox{\boldmath $R$}}
\def\bS{\mbox{\boldmath $S$}}
\def\bT{\mbox{\boldmath $T$}}
\def\bU{\mbox{\boldmath $U$}}
\def\bV{\mbox{\boldmath $V$}}

\def\bX{\mbox{\boldmath $X$}}
\def\bY{\mbox{\boldmath $Y$}}

\def\Id{\mbox{\boldmath $I$}}
\newcommand{\bsigma}{\mbox{\boldmath $\sigma$}}
\newcommand{\bepsilon}{\mbox{\boldmath $\epsilon$}}

\def\f0{\mbox{$\mathbb{O}$}}

\def\fE{\mbox{$\mathbb{E}$}}

\def\fG{\mbox{$\mathbb{G}$}}

\def\fM{\mbox{$\mathbb{M}$}}

\newcommand{\mK}{\ensuremath{\mathcal{K}}}

\def\tr{\mbox{$\mathrm{tr}$}}
\def\det{\mbox{$\mathrm{det}$}}
\def\dev{\mbox{$\mathrm{dev}$}}

\def\exp{\mbox{$\mathrm{exp}$}}
\def\cos{\mbox{$\mathrm{cos}$}}

\def\sgn{\mbox{$\mathrm{sgn}$}}
\def\log{\mbox{$\mathrm{log}$}}

\def\Orth{\mbox{$\mathsf{Orth}$}}
\def\Orth+{\mbox{$\mathsf{Orth^+}$}}
\def\Sym{\mbox{$\mathsf{Sym}$}}
\newcommand{\Reals}{\ensuremath{\mathbf{R}}}

\def\CMAME{{\it Comput.\ Method.\ Appl.\ M.}\ }

\def\IJP{{\it Int.\ J.\ Plasticity}\ }
\def\IJSS{{\it Int.\ J.\ Solids Struct.}\ }

\def\JAM{{\it J.\ Appl.\ Mech.}\ }

\def\JMPS{{\it J.\ Mech.\ Phys.\ Solids}\ }

\begin{document}

\title{An elastoplastic framework for granular materials 
becoming cohesive through mechanical densification. \\
Part II - the formulation of elastoplastic coupling at large strain}

\author{\\Andrea Piccolroaz, Davide Bigoni and Alessandro Gajo \\
Dipartimento di Ingegneria Meccanica e \\ Strutturale,
Universit\`a di Trento, \\ Via Mesiano 77, I-38050 Trento, Italia\\
email: bigoni@ing.unitn.it}

\date{August 2, 2004}

\maketitle

\begin{abstract}
\noindent
The two key phenomena occurring in the process of ceramic powder compaction are the progressive gain in cohesion and 
the increase of elastic stiffness, both related to 
the development of plastic deformation. 
The latter effect is an example of `elastoplastic coupling', in which the plastic flow affects the elastic properties of the 
material, and has been so far considered only within the framework of small strain assumption 
(mainly to describe elastic degradation in rock-like materials), so that it remains completely unexplored for large strain.
Therefore, a new finite strain generalization of elastoplastic coupling theory is given to describe the 
mechanical behaviour of materials evolving from a granular to a dense state. 

The correct account of elastoplastic coupling and of the specific characteristics of materials evolving from a 
loose to a dense state (for instance, nonlinear --or linear-- dependence of the elastic part of the deformation on 
the forming pressure in the granular --or dense-- state) makes the use of existing large strain
formulations awkward, if even possible. Therfore, first, we have resorted to a very general setting allowing general 
transformations between work-conjugate stress and strain measures; second, we have introduced 
the multiplicative decomposition of the deformation gradient and, third, employing 
isotropy and hyperelasticity of elastic response, we have obtained a relation between the Biot stress 
and its `total' and `plastic' work-conjugate strain measure. This is a key result, since it allows an 
immediate achievement of the rate elastoplastic constitutive equations. Knowing the general form of these 
equations, all the specific laws governing the behaviour of ceramic powders are 
finally introduced as generalizations of the small strain counterparts given in Part I of this paper.
\end{abstract}


\thispagestyle{empty} 


\section{Introduction}

Mechanical cold compaction of ceramic powder involves the transition from a granular to a 
dense state. During this process, and strictly related to the development of permanent deformations, 
both the cohesion {\it and elastic stiffness} of the material increase. This occurs also under isostatic 
compression (which does not involve deviatoric strain) and is believed to be related at the microscale 
to the increase of the contact area between the grains (therefore, the effect should not be confused 
with a large strain effect). The increase of elastic stiffness with permanent deformation is a sort of
`inverse damage', which can be described making recourse to the concept of elastoplastic coupling, originarily 
invented to model elastic degradation, and
employed 
in the Part I of this paper to describe the stiffening during plastic deformation of ceramic powder. 
However, a large strain formulation of elastoplastic coupling has never been 
attempted. That this formulation is not trivial can be deduced from the 
fact that elastic characteristics have been assumed to be independent of plastic deformation in 
all elastoplastic models proposed for soils (Borja and Tamagnini, 1998; 
Callari et al. 1998; Rouainia and Muir Wood, 2000; 
Ortiz and Pandolfi, 2004) and in more general contexts (see among others: Simo and Miehe, 1992; Peric et al. 1992; 
Schieck and Stumpf, 1993; Simo and Meschke, 1993; Ibrahimbegovic, 1994).

Since the existing large strain formulations do not appear to be easily 
generalizable to admit a coupling between elastic and plastic deformations, we have recurred to the 
early formulation by Hill and Rice (1973) (see also Hill, 1978; Petryk and Thermann 1985; 
Bigoni, 1996; 2000), 
which (although not explicitely mentioned) has been formulated in such a generality to include coupling.
Following this approach, the level of generality is so high that the following choices are not required: 
stress/strain measures [except that these are work-conjugate (Hill, 1968)], elastic 
and plastic strain decomposition, elastic law, yield function, 
flow and hardening rules. 
After this framework is provided, the multiplicative strain decomposition of Lee (1969) and Willis (1969) is 
introduced. At this point, assuming that the elastic response be hyperelastic and isotropic we have proved that
a general relation exists, in which the Biot stress is related to its work-conjugate 
`total' and `plastic' strain measure (and to a generic set of scalar hidden variables). This achievement turns out to be  
crucial since it allows immediate use of the general formulation previously developed\footnote{The law between
Biot stress and its work-conjugate strain measure could obviously be transformed into different (work-conjugate) 
stress/strain measures, but this would be cumbersome and useless, since the generality of the Hill and Rice (1973) 
approach allow us to use the obtained law directly.}.
Finally, the coupling and hardening laws, the yield function and all constitutive relations provided in Part I of this
paper for the infinitesimal theory are consistently generalized to include large strains.

\section{The skeleton of large strain elastoplasticity} \lb{sega}

\subsection{Some preliminaries on work conjugacy}

A broad constitutive framework for isothermal and time independent large elastoplastic deformations is presented,
based on the concept of work coniugacy in the Hill sense (1968, 1978). In particular, employing 
Ogden's (1984) notation, a pair of symmetric, Lagrangean, stress $\bT^{(m)}$ and strain $\bE^{(m)}$ 
measures\footnote{
The notation $\bT^{m} = \underbrace{\bT\bT...\bT}_{m~\mathrm{times}}$ (or $\bE^m$) should not be confused with 
$\bT^{(m)}$ (or $\bE^{(m)}$).
} 
are work-conjugate when the stress power density per unit 
volume in the reference configuration can be expressed as
\beq
\lb{workconj}
\bT^{(m)} \scalp \dot{\bE}^{(m)} = \bS \scalp \dot{\bF},
\eeq
where a dot over a symbol denotes material time derivative, 
$\bF$ is the deformation gradient and $\bS$ the first Piola-Kirchhoff stress tensor
\beq
\bS = J \bT \bF^{-T} = \bK \bF^{-T},
\eeq
in which $J = \det \bF$ and $\bT$ and $\bK = J\bT$ are the Cauchy and Kirchhoff stresses, respectively.

For integer (positive, null or negative) exponent $m$, we introduce the following Lagrangean strain measures
\beq
\lb{strains}
\left\{
\barr{ll}
\ds
\bE^{(m)} = \frac{1}{m} (\bU^m - \Id), &~~~ \mathrm{if}~~ m \neq 0, \\[5 mm]
\bE^{(0)} = \log\, \bU , &~~~ \mathrm{if}~~ m = 0, 
\earr
\right.
\eeq
where the logarithm of a tensor is defined as in (Ogden, 1984) and 
\beq
\bU = (\bF^T \bF)^{1 / 2} ,
\eeq 
is the right stretch tensor. For a given $m$, $\bE^{(m)}$ is defined by 
(\ref{strains}) and the corresponding work-conjugate stress measure $\bT^{(m)}$ can be 
defined imposing eqn.~(\ref{workconj}). For instance, 
for $m = 2$, the Green-Lagrange strain results from eqn.~(\ref{strains}) and the eqn.~(\ref{workconj}) provides 
for $\bT^{(2)}$ the second Piola-Kirchhoff stress tensors,
\beq
\lb{secondo}
\bE^{(2)} = \frac{1}{2} \left( \bU^2 - \Id \right), ~~~\mathrm{conjugate~to}~~~ 
\bT^{(2)} = \bF^{-1} \bK \bF^{-T}.
\eeq
A conjugate pair of stress and strain that will become useful later is formed by the Biot stress tensor 
$\bT^{(1)}$ and the strain measure $\bE^{(1)}$, defined as
\beq
\lb{biot}
\bE^{(1)} = \bU - \Id, ~~~\mathrm{conjugate~to}~~~ 
\bT^{(1)} = \frac{1}{2} \left( \bT^{(2)} \bU + \bU \bT^{(2)} \right).
\eeq

It is well-known however that it is not always the easy task of the two above examples to obtain the stress 
measure conjugated to a given strain of the form (\ref{strains}). For instance, the conjugate of the logarithmic 
strain $\bE^{(0)}$ has a very complex form (Hoger, 1987), which simplifies to the rotated stress only when the 
two measures result coaxial, namely, 
\beq
\lb{rotato}
\bE^{(0)} = \log\, \bU ~~~\mathrm{conjugate~to}~~~ \bT^{(0)} = \bR^T \bK \bR, 
\eeq
(where $\bR = \bF\bU^{-1}$ is the rotation tensor) if and only if the following coaxiality condition holds true
\beq
\lb{coaxial}
\bE^{(0)} \bT^{(0)} = \bT^{(0)} \bE^{(0)} ~~~\Longleftrightarrow~~~(\log\bV)\bK = \bK\,\log\bV,
\eeq
where $\bV$ is the left stretch tensor, so that $\bF = \bR\bU = \bV\bR$ (note that the above equivalence is an
immediate consequence of the fact that the logarithmic function is isotropic). 
Condition (\ref{coaxial}) is satisfied for isotropic elasticity, but may be not in more general 
contexts, such as for instance elastoplasticity (Sansour, 2001). It may be instructive for subsequent considerations 
to note that, 
when the coaxiality condition (\ref{coaxial}) holds true, the following relation 
\beq
\lb{eule}
\bT^{(0)} \scalp\dot{\bE}^{(0)} = \bK\scalp\left(\log \bV\right)^\cdot
\eeq
can be proved (Ogden, 1982) showing that the Eulerian stress and
strain measures $\bK$ and $\log \bV$ are work-conjugate (Hill, 1968).

\subsection{The basic assumptions of elastoplasticity}

Following Bigoni (2000), inelastic materials are considered that may at any stage of deformation exhibit a purely 
elastic response for appropriate loading. For these materials, elastic response is assumed to be a one-to-one 
relation between a work-conjugate pair $\bT^{(m)}$ and $\bE^{(m)}$
\beq
\lb{elas}
\bT^{(m)} = \hat{\bT}^{(m)} (\bE^{(m)}, \mK),~~~
\bE^{(m)} = \hat{\bE}^{(m)} (\bT^{(m)}, \mK),
\eeq
where $\hat{\bT}^{(m)}$ and $\hat{\bE}^{(m)}$ are functionals depending on the prior history of inelastic deformation 
through the unspecified set $\mK$ of variables of generic tensorial nature (thus embracing scalars, vectors,
second-order tensors and possibly higher-order tensors). For a purely elastic 
deformation rate, $\mK$ remains fixed, so that we have
\beq
\lb{elastico}
\dot{\bT}^{(m)} = \fE [\dot{\bE}^{(m)}], ~~~ 
\dot{\bE}^{(m)} = \fM [\dot{\bT}^{(m)}],
\eeq
where the fourth-order tensors $\fE$ and $\fM$ possess the minor symmetries induced by $\bE$ and $\bT$, while
the major symmetry is not a-priori requested (differently from Hill and Rice, 1973). They are defined as
\beq
\lb{tensori}
\fE (\bE^{(m)},\mK) = \frac{\partial \hat{\bT}^{(m)}}{\partial \bE^{(m)}},~~~
\fM (\bT^{(m)},\mK) = \frac{\partial \hat{\bE}^{(m)}}{\partial \bT^{(m)}}.
\eeq
Tensors (\ref{tensori}) obviously satisfy
\beq
\fE = \fM^{-1}.
\eeq
For an increment involving elastic and inelastic strain rates, we may write
\beq
\lb{elastoplastico}
\dot{\bT}^{(m)} = \fE[\dot{\bE}^{(m)}] - \dot{\Lambda} \fE[\bP], ~~~ 
\dot{\bE}^{(m)} = \fM[\dot{\bT}^{(m)}] + \dot{\Lambda} \bP,
\eeq
where $\bP \in \Sym$ sets the `direction' (or the `mode') of the irreversible deformation, which is given by
\beq
\lb{flow}
\dot{\Lambda} \bP = 
- \fE^{-1} \frac{\partial \hat{\bT}^{(m)}}{\partial \mK}[\dot{\mK}] =
\frac{\partial \hat{\bE}^{(m)}}{\partial \mK}[\dot{\mK}] .
\eeq
The scalar $\dot{\Lambda} \geq 0$ appearing in eqn. (\ref{elastoplastico}) is 
the plastic multiplier and vanishes for purely elastic response, namely, when $\dot{\mK} = 0$. 

A yield 
surface is assumed at each $\mK$, which may be alternatively expressed in stress and strain spaces as 
\beq
f_{\mbox{\boldmath $\scriptstyle T$}^{(m)}} (\bT^{(m)},\mK) \leq 0 ~~\mbox{or as}~~ 
f_{\mbox{\boldmath $\scriptstyle E$}^{(m)}} (\bE^{(m)},\mK) \leq 0,
\eeq
thus defining regions of the $\bT^{(m)}$ or 
$\bE^{(m)}$ space, respectively, within which the response is elastic. 

\subsection{Direct rate constitutive equations}

Prager's consistency condition requires 
\beq
\dot{f}_{\mbox{\boldmath $\scriptstyle T$}^{(m)}} = \dot{f}_{\mbox{\boldmath $\scriptstyle E$}^{(m)}} = 0,
\eeq
when 
inelastic strain rate is different from zero. As a consequence, employing the stress space representation, the 
elastoplastic incremental constitutive equations can be written as
\beq
\lb{ep}
\dot{\bT}^{(m)} =
\left\{ 
\barr{ll}
\ds
\fE[\dot{\bE}^{(m)}] - \frac{1}{g} <\bQ \scalp \fE[\dot{\bE}^{(m)}]> \fE[\bP] & ~~\mathrm{if}~ 
f_{\mbox{\boldmath $\scriptstyle T$}^{(m)}} (\bT^{(m)}, \mK) = 0, \\[5mm]
\fE[\dot{\bE}^{(m)}] & ~~\mathrm{if}~ f_{\mbox{\boldmath $\scriptstyle T$}^{(m)}} (\bT^{(m)}, \mK) < 0,
\earr 
\right .
\eeq
where the operator $<\cdot>$ denotes the Macaulay brackets, i.e.\ $\forall\ \alpha \in \Reals$, 
$<\alpha> = (\alpha+|\alpha|)/2$. 
Moreover, the symmetric second-order tensor 
\beq
\bQ = \frac{\partial f_{\mbox{\boldmath $\scriptstyle T$}^{(m)}}}{\partial \bT^{(m)}} ,
\eeq
is the yield 
function gradient and the plastic modulus
\beq
\lb{plasticm}
g = h + \bQ \scalp \fE[\bP],
\eeq
is assumed to be strictly positive (a negative plastic modulus would correspond to a so-called locking material, 
not considered here). In the Hill (1967) notation, the hardening modulus $h$ in (\ref{plasticm}) describes:
\begin{itemize}
\item {\it hardening} when positive, 
\item {\it softening} when negative, 
\item {\it perfect plasticity} when null. 
\end{itemize}
The hardening modulus is defined as
\beq
\lb{hardm}
\dot{\Lambda} h = - \frac{\partial f_{\mbox{\boldmath $\scriptstyle T$}^{(m)}}}{\partial \mK} \scalp \dot{\mK},
\eeq
and, as Hill (1967) remarks, hardening and softening are not measure-invariant concepts, in the sense that $h$ 
depends on the choice of $\bT^{(m)}$ and $\bE^{(m)}$. Therefore, the above nomenclature is, to some extent, arbitrary. 
Moreover, we remark that, in addition to $h$, also $\bQ$, $\bP$ and $\fE$ are measure-dependent. On the contrary, 
the plastic modulus $g$ can be shown to be measure-independent (Hill, 1967; Petryk, 2000). Note also that all 
quantities appearing in the rate equations (\ref{ep}) fully depend on the entire path of deformation reckoned 
from some ground state.

\subsection{Inverse rate constitutive equations}

Under the assumption of positive hardening, the rate equations (\ref{ep}) can be inverted to relate
the material derivative of the strain measure to the material derivative of the work-conjugate stress.
In particular, taking the scalar product of the first equation in (\ref{ep}) with $\bQ$ gives
\beq
\lb{pss}
\bQ \scalp \dot{\bT}^{(m)} = \bQ \scalp \fE[\dot{\bE}^{(m)}] - \frac{\bQ \scalp \fE[\bP]}{g}
<\bQ \scalp \fE[ \dot{\bE}^{(m)}]>.
\eeq
In the case when $h > 0$, we note that
$$
\sgn (\bQ \scalp \fE[\dot{\bE}^{(m)}]) = \sgn (\bQ \scalp \dot{\bT}^{(m)}).
$$ 
Therefore, assuming positive hardening, $h > 0$, and using (\ref{pss}), we obtain the inverse constitutive equations
\beq
\lb{epinv}
\dot{\bE}^{(m)} =
\left\{ 
\barr{ll}
\ds
\fM[\dot{\bT}^{(m)}] + \frac{1}{h} <\bQ \scalp \dot{\bT}^{(m)}> \bP & ~~~\mathrm{if}~~ 
f_{\mbox{\boldmath $\scriptstyle T$}^{(m)}} (\bT^{(m)}, \mK) = 0, \\[5mm]
\fM[\dot{\bT}^{(m)}] & ~~~\mathrm{if}~~ f_{\mbox{\boldmath $\scriptstyle T$}^{(m)}} (\bT^{(m)}, \mK) < 0,
\earr 
\right .
\eeq

The rate constitutive equations (\ref{ep}) or (\ref{epinv}) represent a broad constitutive framework, within which 
\begin{quote}
all possible choices of $\bT^{(m)}$ and $\bE^{(m)}$ 
are equivalent and the requirement of material 
frame indifference (Truesdell and Noll, 1965) is never violated.
\end{quote}
Moreover, the framework is so general that it does not imply any particular choice of 
\begin{itemize}
\item elastic and plastic strain decomposition,
\item hypo- or hyper-  elastic law,
\item yield function, flow and hardening rules.
\end{itemize}
It is however clear that in order to set up the constitutive modelling of a particular material, we need to 
introduce specific laws. This objective will be pursued in three steps of deceasing generality in the following: 
first, we will 
introduce the multiplicative elastic and plastic strain decomposition and 
requirement of objectivity and isotropy of the elastic constitutive law; 
second, a form of elastic constitutive equation will be proposed, depending on plastic deformation and thus capable 
of describing the elastic behaviour of granular and dense materials; third, yield 
function and hardening laws 
are introduced as simple generalizations of the rules formulated in Part I of this paper under the 
small strain assumption.

\section{The multiplicative decomposition and the elastic law}
\begin{figure}[!htb]
\begin{center}
\vspace*{3mm}
\includegraphics[width=10cm]{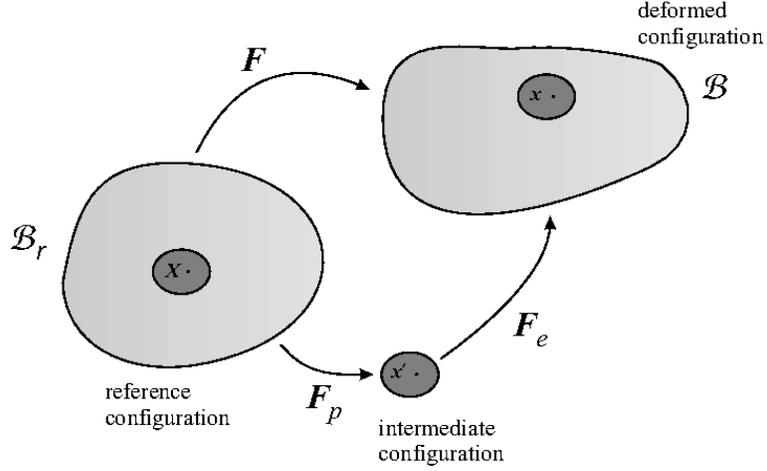}
\caption{\footnotesize Reference, current and intermediate configurations.}
\label{fig28}
\end{center}
\end{figure}

The multiplicative decomposition of deformation gradient $\bF$ into elastic $\bF_e$ and plastic $\bF_p$ 
components introduced by Lee (1969) and Willis (1969) is adopted (Fig.~\ref{fig28}).
\beq
\lb{decomp}
\bF = \bF_e \bF_p.
\eeq
According to eqn.~(\ref{decomp}), using the left polar decomposition $\bF = \bV \bR$, we introduce the elastic 
and plastic left stretch and rotation tensors $\bV_e$, $\bV_p$, $\bR_e$ and $\bR_p$, satisfying
\beq
\bF_e = \bV_e \bR_e, ~~~\bF_p = \bV_p \bR_p,
\eeq
while using the right polar decomposition $\bF = \bR \bU$, the right elastic and plastic stretch tensors $\bU_e$ 
and $\bU_p$ are defined so that they satisfy
\beq
\bF_e = \bR_e \bU_e, ~~~\bF_p = \bR_p \bU_p.
\eeq

A crucial expedient, employed also by Ortiz and Pandolfi (2004), to describe the behaviour of granular materials is to refer to the logarithmic strains 
defined as
\beq
\lb{log1}
\bepsilon = \log\, \bV, ~~~
\bepsilon_e = \log\, \bV_e, ~~~
\bepsilon_p = \log\, \bV_p, 
\eeq
and 
\beq
\lb{log2}
\bE^{(0)} = \log\, \bU, ~~~
\bE^{(0)}_e = \log\, \bU_e, ~~~
\bE^{(0)}_p = \log\, \bU_p.
\eeq
The interest in employing definitions (\ref{log1}) and (\ref{log2}) is that these allow a decoupling between 
the volumetric logarithmic elastic and plastic deformations, namely
\beq
\tr\, \bepsilon = \tr\, \bepsilon_e + \tr\, \bepsilon_p = 
\tr\, \bE^{(0)} = \tr\, \bE^{(0)}_e + \tr\, \bE^{(0)}_p,
\eeq
which, employing the usual definition of $J$ and noting the property (for every symmetric tensor $\bA$)
\beq
\tr(\log\, \bA) = \log(\det\, \bA),
\eeq
can be written as
\beq
\log\, J = \log\, J_e + \log\, J_p.
\eeq

\section{Objectivity and isotropy of elastic response}

We refer now to an isotropic elastic law relating the Kirchhoff stress $\bK$ to the elastic deformation 
gradient $\bF_e$ in the generic form 
\beq
\lb{kirk}
\bK = \hat{\bK}(\bF_e,k_i),
\eeq
where {\it function $\hat{\bK}$ may depend also on generic plastic scalar variables $k_i$, assumed invariant with 
respect to every symmetry group of the material and change in observer}. 

In general, the elastic response must be objective, but in addition we assume for simplicity that the elastic response 
be isotropic. The latter requirement implies the coaxiality condition (\ref{coaxial}), a requisite 
more important than it may appear, since it ensures the work coniugacy (\ref{rotato}) and its Eulerian counterpart.
Therefore, function $\hat{\bK}$ is assumed to satisfy:
\begin{description}
\item[P1.] The objectivity requirement 
\beq
\hat{\bK}(\bF_e,k_i) = \bR^T \hat{\bK}(\bR \bF_e,k_i) \bR, ~~~\forall \bR \in \Orth+.
\eeq
\item[P2.] The isotropy requirement 
\beq
\lb{iso}
\hat{\bK}(\bF_e,k_i) = \hat{\bK}(\bF_e \bR,k_i), ~~~\forall \bR \in \Orth+.
\eeq
\end{description}
Though the generalization of the formulation to anisotropy of the elastic response is de\-fi\-ni\-tely important to
capture certain experimental evidence connected to the development of 
various form of instabilities [see Gajo et al. (2004) for a discussion relative to the infinitesimal theory
in the context of granular media], we note that this generalization is for the moment lacking for 
granular material subject to large strains, even in the relatively simple setting in which 
cohesion and coupling are neglected.

As a consequence of isotropy, eqn. (\ref{iso}), the rotation in the left polar decomposition does not alter the values 
of function $\hat{\bK}$, 
\beq
\hat{\bK}(\bF_e,k_i) = \hat{\bK}(\bV_e,k_i),
\eeq
so that function $\hat{\bK}$ depends only on the elastic left stretch tensor. 
Noting the identity
\beq
\bF_e = \bR \bU \bU_p^{-1} \bR_p^T,
\eeq
isotropy and objectivity allow us to introduce the following transformations
\beq
\bK = \hat{\bK}(\bR \bU \bU_p^{-1} \bR_p^T,k_i)  = \hat{\bK}(\bR \bU \bU_p^{-1},k_i) = 
\bR \hat{\bK}(\bU \bU_p^{-1},k_i) \bR^T,
\eeq
so that we get
\beq
\bK = \hat{\bK}(\bF_e,k_i) = \bR \hat{\bK}(\bU \bU_p^{-1},k_i) \bR^T.
\eeq
Employing now the rotated stress $\bR^T\bK\bR$, the constitutive law (\ref{kirk}) can be written in the form
\beq
\lb{kirk2}
\bR^T\bK\bR = \hat{\bK}(\bU \bU_p^{-1},k_i),
\eeq
relating the rotated stress to the total and plastic right stretch tensors.

Now, the rotated stress is related to the Biot stress through (Ogden, 1984) 
\beq
\bT^{(1)} = \frac{1}{2} \left(\bU^{-1} \bR^T \bK \bR  + \bR^T \bK \bR \bU^{-1} \right),
\eeq
so that in conclusion we obtain the elastic constitutive law in the form
\beq
\lb{biot2}
\bT^{(1)} = \frac{1}{2} \left(\bU^{-1} \hat{\bK}(\bU \bU_p^{-1},k_i) 
+ \hat{\bK}(\bU \bU_p^{-1},k_i) \bU^{-1} \right),
\eeq
relating the Biot stress to the global and plastic right stretch tensors.

Since eqn. (\ref{biot}) shows that tensor $\bE^{(1)}$ is the right stretch tensor $\bU$ with the identity subtracted, 
\begin{quote}
eqn.~(\ref{biot2}) expresses a relation between the two work-conjugate measures 
$\bT^{(1)}$ and $\bE^{(1)}$ of the type (\ref{elas}), in which the set $\mK$ is now including
$\bU_p^{-1}$ and the scalars $k_i$. 
\end{quote}

\section{Formulation of the rate model}

Until this point, all the equations are at a high level of generality; now,
to develop the model for powder densification, further specific laws are
introduced, including the particular hyperelastic-plastic coupling rule, 
yield function and hardening laws. Since these laws are essentially extensions of those
already employed in the small strain formulation, details on physical motivations determining the
specific choices will be omitted for conciseness.

\subsection{Elastoplastic coupling}

The elastic properties of granular materials can be described by a hyperelastic nonlinear law providing a 
generalization to finite strains of the corresponding equation introduced in Part I, Section 2.5. 
This generalization is represented by the following potential
\beqar
\lb{pot}
\lefteqn{\phi(\bepsilon_e,\bepsilon_p) = 
- \frac{\mu}{3} (\tr\, \bepsilon_e)^2 + c\, \tr\, \bepsilon_e} \\ 
& & ~~~~
+ (p_0 + c) \left[ \left( d - \frac{1}{d} \right) \frac{(\tr\, \bepsilon_e)^2}{2 \tilde{\kappa}}
+ d^{1/n} \tilde{\kappa}\, \exp \left( - \frac{\tr\, \bepsilon_e}{d^{1/n} \tilde{\kappa}} \right) \right]
+ \mu\, \bepsilon_e \scalp \bepsilon_e , \nonumber
\eeqar
where $\tilde{\kappa}$ is the elastic logarithmic bulk modulus, $p_0$ is the initial confining pressure, 
$c$, $d$, and $\mu$ are scalar parameters depending on the 
volumetric plastic strain $\tr\, \bepsilon_p = \tr\, \bE^{(0)}_p$, providing the elastoplastic coupling.
The dependence of parameters $c$, $d$, and $\mu$ on the volumetric plastic strain is made explicit by the 
following equations
\beq
\lb{mudici}
\barr{l}
\ds \mu = \mu_0 + c \left( d - \frac{1}{d} \right) \mu_1, ~~~ d = 1 + B <p_c - p_{cb}>, \\[5mm]
c = c_{\infty} \left[ 1- \exp \left(-\Gamma <p_c - p_{cb}> \right) \right],
\earr
\eeq
and
\beq
\lb{doppioexp2}
\exp \left(\tr\, \bE_{p}^{(0)} \right) - 1 = 
- \tilde{a}_1\, \exp \left( -\frac{\Lambda_1}{p_c} \right) 
- \tilde{a}_2\, \exp \left( -\frac{\Lambda_2}{p_c} \right),
\eeq
where $\mu_0$, $\mu_1$, $B$, $p_{cb}$, $c_{\infty}$, $\Gamma$, $\tilde{a}_1$, $\tilde{a}_2$, $\Lambda_1$, 
and $\Lambda_2$ are positive material constants. 
It can be noted that eqns. (\ref{mudici}) do not include `geometrical terms' and thus 
coincide with those of the small strain formulation, whereas eqn. (\ref{doppioexp2}) has been consistently 
generalized.

The potential (\ref{pot}) represents a isotropic function of the logarithmic strain and can be
written as the sum of a volumetric and a deviatoric component. The deviatoric potential coincides
with that employed by Ortiz and Pandolfi (2004) in the special case of null cohesion $c=0$ and null coupling 
$d=1$.

The Kirchhoff stress can be obtained from the potential (\ref{pot}) as
\beq
\lb{ki}
\bK = \deriv{\phi}{\bepsilon_e}, 
\eeq
so that it results in the form
\beq
\lb{kirk3}
\bK = 
\left\{ -\frac{2}{3} \mu\, \tr\, \bepsilon_{e} + c  
 + (p_0 + c) \left[ \left( d - \frac{1}{d} \right) \frac{\tr\, \bepsilon_{e}}{\tilde{\kappa}}
- \exp \left( -\frac{\tr\, \bepsilon_{e}}{d^{1/n} \tilde{\kappa}} \right) \right] \right\} \Id 
+ 2 \mu\, \bepsilon_{e}.
\eeq
Eqn. (\ref{kirk3}) implies that the Kirchhoff stress and the logarithmic strain are coaxial, so that these
become in the present context work-conjugate stress and strain measures, eqn. (\ref{eule}). 

The elastic constitutive law (\ref{kirk3}) can be written in the form (\ref{kirk}) with 
\beqar
\lb{kirk4}
\lefteqn{\hat{\bK}(\bF_e,k_i) = \left\{ -\frac{1}{3} \mu\, \tr\, \log\, \bF_e \bF_e^T + c \right.} \\ 
& & 
\left. + (p_0 + c) \left[ \left( d - \frac{1}{d} \right) \frac{\tr\, \log\, \bF_e \bF_e^T}{2 \tilde{\kappa}}
- \exp 
\left( -\frac{\tr\, \log\, \bF_e \bF_e^T}{2 d^{1/n} \tilde{\kappa}} \right) \right] \right\} \Id \nonumber \\  
& &
+ \mu\, \log\, \bF_e \bF_e^T. \nonumber
\eeqar
where the set $k_i$ is now including $c$, $d$, and $\mu$. 
Obviously, eqn.~(\ref{kirk4}) can be expressed in the 
form (\ref{biot2}), not reported for conciseness. 

We are now in a position to write down the fourth-order 
elastic tensor $\fE$ defined by eqn.~$(\ref{tensori})_1$. This takes the form
\beqar
\lb{eltens}
\lefteqn{
\fE = -\frac{1}{2} \left( \bU^{-1} \bob \hat{\bK} \bU^{-1} + \hat{\bK} \bU^{-1} \bob \bU^{-1} \right)} \\
& & 
+\frac{1}{2} \left( \deriv{\bU^{-1} \hat{\bK}(\bX,k_i)}{\bX} + 
\deriv{\hat{\bK}(\bX,k_i) \bU^{-1}}{\bX} \right)_
{\mbox{\boldmath $\scriptstyle X$} = \mbox{\boldmath $\scriptstyle U$} \mbox{\boldmath $\scriptstyle U$}_p^{-1}} 
\left( \Id \bob \bU_p^{-1} \right), \nonumber
\eeqar
so that, employing eqn. (\ref{kirk4}), we obtain
\beqar
\lefteqn{
\deriv{\hat{\bK}(\bX,k_i)}{\bX} = } \\
& & 
\left\{ \left[ -\frac{\mu}{3} + K_t(\bX) \right] \Id \otimes \Id + \mu \Id \bob \Id \right\}
\left( \frac{\partial\, \log\, \bY}{\partial \bY} \right)_
{\mbox{\boldmath $\scriptstyle Y$} = \mbox{\boldmath $\scriptstyle X$} \mbox{\boldmath $\scriptstyle X$}^T}
\left( \Id \sot \bX + \bX \sop \Id \right), \nonumber
\eeqar
where
\beq
K_t(\bX) = 
\frac{p_0 + c}{2 \tilde{\kappa}} \left[ d - \frac{1}{d} + d^{-1/n} 
\exp \left( -\frac{\tr\, \log\, \bX \bX^T}{2 d^{1/n} \tilde{\kappa}} \right) \right].
\eeq
and 
\beq
\lb{logy}
\frac{\partial\, \log\, \bY}{\partial \bY} = 
\sum_{n = 1}^{\infty} \frac{(-1)^{n + 1}}{n} 
\sum_{r = 0}^{n-1} \left( \bY - \Id \right)^r \bob \left( \bY - \Id \right)^{n-1-r},
\eeq
a formula that can be found in (Hoger, 1987) or deduced from (Itskov and Aksel, 2002).
Note that four tensorial products between second-order tensors $\bA$ and $\bB$ have been employed, which can be 
defined, with reference to every tensor $\bC$, as
\beq
\barr{ll}
(\bA \otimes \bB)[\bC] = (\bC \scalp \bB^T) \bA, &~~~ 
\ds (\bA \bob \bB)[\bC] = \frac{1}{2} \bA \left( \bC + \bC^T \right) \bB^T, \\[5mm]
(\bA \sot \bB)[\bC] = \bA \bC \bB^T, &~~~
(\bA \sop \bB)[\bC] = \bA \bC^T \bB^T,
\earr
\eeq
so that the following property holds
\beq
\bob = \frac{1}{2} \left( \sot + \sop \right).
\eeq

\subsection{The irreversible strain rate}

Eqn.~(\ref{flow})$_1$ defines the irreversible strain rate, which, accordingly, may be calculated taking the 
derivatives of eqn. (\ref{biot2}) in which $\hat{\bK}$ is given by eqn. (\ref{kirk4}). We obtain
\beq
\lb{pippo}
\dot{\Lambda} \bP = \fG \left[ \dot{\bE}_p^{(0)} \right],
\eeq
where tensor $\fG$, assumed positive definite, is given by
\beqar
\lb{tensgi}
\fG & = &
- \frac{1}{2} \xi_2\, \fE^{-1} 
\left[ \left( \bU^{-1} \deriv{\hat{\bK}}{c} + \deriv{\hat{\bK}}{c} \bU^{-1} \right) \otimes \Id \right] \nonumber \\[3mm]
& &
- \frac{1}{2} \xi_3\, \fE^{-1} 
\left[ \left( \bU^{-1} \deriv{\hat{\bK}}{d} + \deriv{\hat{\bK}}{d} \bU^{-1} \right) \otimes \Id \right] \nonumber \\[3mm]
& &
- \frac{1}{2} \xi_4\, \fE^{-1} 
\left[ \left( \bU^{-1} \deriv{\hat{\bK}}{\mu} + \deriv{\hat{\bK}}{\mu} \bU^{-1} \right) \otimes \Id \right] \\[3mm]
& & 
+ \frac{1}{2}\, \fE^{-1} 
\left( \deriv{\bU^{-1} \hat{\bK}(\bX)}{\bX} + \deriv{\hat{\bK}(\bX) \bU^{-1}}{\bX} \right)_
{\mbox{\boldmath $\scriptstyle X$} = \mbox{\boldmath $\scriptstyle U$} \mbox{\boldmath $\scriptstyle U$}_p^{-1}} \cdot \nonumber \\[3mm]
& & 
\cdot \left( \bU \bU_p^{-1} \bob \bU_p^{-1} \right) \frac{\partial\, \exp\, \bE_p^{(0)}}{\partial \bE_p^{(0)}}, \nonumber
\eeqar
in which 
\beqar
\lefteqn{
\deriv{\hat{\bK}}{c} = } \\ 
& & 
\left\{
1 + \left[ \left( d - \frac{1}{d} \right) \frac{\tr\, \log\, \bU \bU_p^{-2} \bU}{2 \tilde{\kappa}} 
- \exp \left( - \frac{\tr\, \log\, \bU \bU_p^{-2} \bU}{2 d^{1/n} \tilde{\kappa}}  \right) \right]
\right\} \Id, \nonumber
\eeqar
\beqar
\lefteqn{
\deriv{\hat{\bK}}{d} = 
\left\{
\frac{(p_0 + c)\, \tr\, \log\, \bU \bU_p^{-2} \bU}{2 \tilde{\kappa}} 
\left[ 1 + \frac{1}{d^2} \right. \right.} \\
& & ~~~~~~~~~~~~~~~~~~~~
\left. \left. - \frac{1}{2 n d^{1+1/n}}\, 
\exp \left( - \frac{\tr\, \log\, \bU \bU_p^{-2} \bU}{2 d^{1/n} \tilde{\kappa}}  \right) \right]
\right\} \Id, \nonumber
\eeqar
\beq
\deriv{\hat{\bK}}{\mu} = 
\left( -\frac{1}{3}\, \tr\, \log\, \bU \bU_p^{-2} \bU \right) \Id + \log\, \bU \bU_p^{-2} \bU,
\eeq
and
\beq
\xi_2 =  
- \frac{c_{\infty} \Gamma\, H(p_c - p_{cb})\, 
\exp \left[-\Gamma ( p_c - p_{cb}) \right] p_c^2\, \exp(\tr\, \bE_p^{(0)})}
{\tilde{a}_1 \Lambda_1\, \exp \left(\ds{- \frac{\Lambda_1}{p_c}} \right)
+ \tilde{a}_2 \Lambda_2\, \exp \left(\ds{- \frac{\Lambda_2}{p_c}} \right)}, 
\eeq
\beq
\xi_3 = 
- \frac{B\, H(p_c - p_{cb})\, p_c^2\, \exp(\tr\, \bE_p^{(0)})}
{\tilde{a}_1 \Lambda_1\, \exp \left(\ds{- \frac{\Lambda_1}{p_c}} \right)
+ \tilde{a}_2 \Lambda_2\, \exp \left(\ds{- \frac{\Lambda_2}{p_c}} \right)},
\eeq
\beq
\xi_4 = 
\left( d - \frac{1}{d} \right) \mu_1 \xi_2 + c \left( 1 + \frac{1}{d^2} \right) \mu_1 \xi_3. 
\eeq
Note that the exponential of a tensor [see e.g. (Itskov and Aksel, 2002) for the definition] has been introduced 
in eqn.~(\ref{tensgi}), together with its gradient, defined as
\beq
\lb{exp}
\frac{\partial\, \exp\, \bE_p^{(0)}}{\partial \bE_p^{(0)}} = 
\sum_{n = 1}^{\infty} \frac{1}{n!} \sum_{r = 0}^{n - 1} \left( \bE_p^{(0)} \right)^r \bob 
\left( \bE_p^{(0)} \right)^{n - 1 - r},
\eeq
a formula given by Itskov and Aksel (2002).

\subsection{The yield function}

In the absence of {\it ad hoc} experimental results, we employ for simplicity a yield function with same form 
adopted for the infinitesimal theory (see Part I of this paper and Bigoni and Piccolroaz, 2004), where now 
the Cauchy stress is replaced by the Biot 
stress $\bT^{(1)}$. This can be pursued re--defining the invariants $p$, $q$ and $\theta$ in terms of Biot stress 
\beq
\lb{invar}
p = -\frac{\tr\, \bT^{(1)}}{3},~~~ q = \sqrt{3 J_2},~~~ 
\theta = \frac{1}{3} \cos^{-1} \left( \frac{3 \sqrt{3}}{2} \frac{J_3}{J_2^{3/2}} \right),
\eeq
where $\theta \in [0, \pi/3]$ and 
\beq
\lb{invar2}
\barr{l}
\ds
J_2 = \frac{1}{2}\, \dev\, \bT^{(1)} \scalp \dev\, \bT^{(1)},~~~ 
J_3 = \frac{1}{3}\, \tr\, \left( \dev\, \bT^{(1)} \right)^3,~~~ \\[3mm]
\ds
\dev\, \bT^{(1)} = \bT^{(1)} - \frac{\tr\, \bT^{(1)}}{3} \Id .
\earr
\eeq
As a consequence, the yield function takes a form of the type 
$f_{\mbox{\boldmath $\scriptstyle T$}^{(1)}}(\bT^{(1)},\mK) \leq 0$, namely,
\beq
\lb{wlf}
F(\bT^{(1)}, p_c, c) = f(p, p_c, c) + \frac{q}{g(\theta)},
\eeq
where $p_c$ and $c$ are the parameters governing the change in shape of the yield surface caused by the 
hardening (as in the infinitesimal theory), and 
\beq
\lb{puzza}
f(p, p_c, c) = 
\left\{
\barr{ll}
- M p_c \sqrt{\left( \Phi - \Phi^m \right) \left[ 2 (1 - \alpha) \Phi + \alpha \right]}
&~~~\mathrm{if}\ \Phi \in [0,1], \\[5mm]
+ \infty 
&~~~\mathrm{if}\ \Phi \notin [0,1],
\earr
\right.
\eeq
in which 
\beq
\Phi = \frac{p + c(\tr\, \bE^{(0)}_p)}{p_c(\tr\, \bE^{(0)}_p) + c(\tr\, \bE^{(0)}_p)}
\eeq
and 
\beq
\lb{giditeta}
g(\theta) = \frac{1}{\ds \cos{\left[ \beta \frac{\pi}{6} - 
\frac{1}{3}\, \cos^{-1} \left( \gamma\, \cos\, 3 \theta \right) \right]}}.
\eeq
Note that $M$, $m$, $\alpha$, $\beta$, and $\gamma$ are material parameters with the same meaning 
as in the infinitesimal theory (already described in 
Part. I of this paper and by Bigoni and Piccolroaz, 2004).

The yield function gradient 
\beq
\lb{cazzone}
\bQ = \frac{\partial F(\bT^{(1)}, p_c, c)}{\partial \bT^{(1)}},
\eeq
can be obtained directly from the yield function (\ref{puzza}) or 
from the corresponding 
equations in Part I of this paper (Appendix A), with $\bT^{(1)}$ replacing $\bsigma$.

\subsection{Flow rule and hardening modulus}

The flow mode tensor $\bP$ is postulated in the form
\beq
\lb{modi}
\bP = \bQ - \frac{\tr\, \bQ}{3} \epsilon\, (1 - \Phi)\, \Id, 
\eeq
where $0 \leq \epsilon \leq 1$  is a nonassociativity parameter, null for associative flow rule.

The hardening modulus can be calculated from the definition (\ref{hardm}) in the form
\beq
\lb{pluto}
\dot{\Lambda} h = - \left( \deriv{F}{p_c} \dot{p}_c + \deriv{F}{c} \dot{c} \right),
\eeq
where
\beqar
\lb{topolino1}
\lefteqn{\deriv{F}{p_c} = - M \sqrt{\left( \Phi - \Phi^m \right)[ 2 (1 - \alpha) \Phi + \alpha]}} \\
& & 
+ M \frac{p_c (p + c)}{(p_c + c)^2}
\frac{\left( 1 - m \Phi^{m-1} \right) \left[ 2 (1 - \alpha) \Phi + \alpha \right] + 2 (1 - \alpha) 
\left( \Phi - \Phi^m \right)}
{2 \sqrt{\left( \Phi - \Phi^m \right) \left[ 2 (1 - \alpha) \Phi + \alpha \right]}}, \nonumber
\eeqar
and
\beqar
\lb{topolino2}
\lefteqn{\deriv{F}{c} = 
- M \frac{p_c (p_c - p)}{(p_c + c)^2} \cdot} \\
& & ~~~~~~~~~~
\cdot \frac{\left( 1 - m \Phi^{m-1} \right) \left[ 2 (1 - \alpha) \Phi + \alpha \right] 
+ 2 (1 - \alpha) \left( \Phi - \Phi^m \right)}
{2 \sqrt{\left( \Phi - \Phi^m \right) \left[ 2 (1 - \alpha) \Phi + \alpha \right]}}, \nonumber
\eeqar
in which 
\beq
\lb{paperino1}
\dot{p}_c = 
- \frac{p_c^2\, \exp(\tr\, \bE_p^{(0)})}{\tilde{a}_1 \Lambda_1\, \exp \left(\ds - \frac{\Lambda_1}{p_c} \right)
+ \tilde{a}_2 \Lambda_2\, \exp \left(\ds - \frac{\Lambda_2}{p_c} \right)}\, \tr\, \dot{\bE}_p^{(0)},
\eeq
and
\beq
\lb{paperino2}
\dot{c} =  
c_{\infty} \Gamma\, H(p_c - p_{cb})\, \exp \left[ - \Gamma (p_c - p_{cb}) \right] \dot{p}_c.
\eeq
Parameters $\Lambda_1$, $\Lambda_2$, $\tilde{a}_1$, $\tilde{a}_2$, $c_\infty$, $\Gamma$, and $p_{cb}$ have been 
introduced and motivated in Part I of this paper.

\subsection{The rate constitutive equations for the compaction model}

The elastoplastic incremental constitutive equations (\ref{ep}), written in terms of Biot stress $\bT^{(1)}$ 
and conjugate strain $\bE^{(1)} = \bU - \Id$, take the form
\beq
\dot{\bT}^{(1)} =
\left\{ 
\barr{ll}
\ds
\fE[\dot{\bE}^{(1)}] - \frac{1}{g} <\bQ \scalp \fE[\dot{\bE}^{(1)}]> \fE[\bP] & ~~\mathrm{if}~ 
F(\bT^{(1)}, p_c, c) = 0, \\[5mm]
\fE[\dot{\bE}^{(1)}] & ~~\mathrm{if}~ F(\bT^{(1)}, p_c, c) < 0,
\earr 
\right . 
\eeq   
where the elastic tensor $\fE$ is given by eqns.~(\ref{eltens})--(\ref{logy}), the yield function 
$F(\bT^{(1)}, p_c, c)$ by  
eqns.~(\ref{wlf})--(\ref{giditeta}) and the yield function gradient $\bQ$ and flow mode tensor $\bP$ 
by eqns. (\ref{cazzone}) and (\ref{modi}). The plastic modulus $g$ is provided by eqn. (\ref{plasticm}), where $h$, the hardening modulus,
is obtained substituting eqn.~(\ref{pippo}) into  
eqns.~(\ref{paperino1})--(\ref{paperino2}) and eqn.~(\ref{pluto}), thus yielding
\beq
h = - \left( \deriv{F}{p_c} \bar{p}_c + \deriv{F}{c} \bar{c} \right),
\eeq
where $\partial F / \partial p_c$ and $\partial F / \partial c$ are specified by 
eqns.~(\ref{topolino1})--(\ref{topolino2}) and 
\beq
\barr{l}
\ds
\bar{p}_c = 
- \frac{p_c^2\, \exp(\tr\, \bE_p^{(0)})}{\tilde{a}_1 \Lambda_1\, \exp \left(\ds - \frac{\Lambda_1}{p_c} \right)
+ \tilde{a}_2 \Lambda_2\, \exp \left(\ds - \frac{\Lambda_2}{p_c} \right)}\, \tr\, \fG^{-1}[\bP], \\[12mm]
\bar{c} =  
c_{\infty} \Gamma\, H(p_c - p_{cb})\, \exp\, \left[ - \Gamma (p_c - p_{cb}) \right] \bar{p}_c,
\earr
\eeq
in which tensor $\fG$ is explicited by eqns.~(\ref{tensgi})--(\ref{exp}).

\section{Conclusions}

A new, consistent generalization to large strains of elastoplasticity theory with coupling 
between elastic and plastic properties 
has been given, based on 
work-conjugate variables and isotropy of the elastic response. This has permitted the extension to large strains 
of the model introduced in Part I of this paper,
to describe granular materials becoming cohesive during mechanical, cold densification.
Therefore, the model allows the simulation of forming processes of green bodies from ceramic powders, including 
situations in which large strains are involved.

\vspace*{3mm}

\vspace*{3mm}
\noindent
{\sl Acknowledgments }

\noindent
Financial support of 
MIUR-COFIN 2003 `Fenomeni di degrado meccanico di interfacce
in sistemi strutturali: applicazioni in Ingegneria Civile ed a campi di ricerca
emergenti' is gratefully acknowledged.


{
\singlespace

}


\begin{thebibliography}{}

\bibitem{big}
Bigoni, D. (1996) 
On smooth bifurcations in non-associative elastoplasticity 
\JMPS 44, 1337-1351.

\bibitem{bigo}
Bigoni, D., (2000) 
Bifurcation and instability of non�associative elastic�plastic solids. 
In CISM Lecture Notes No. 414 "Material Instabilities in Elastic and Plastic Solids", 
(editor H. Petryk) Springer-Verlag, Wien�New York, pp. 1-52.

\bibitem{bigopicc}
Bigoni, D., Piccolroaz, A. (2004) 
Yield criteria for quasibrittle and frictional materials. 
\IJSS 41, 2855-2878.

\bibitem{borj}
Borja, R.I. and Tamagnini, C. (1998) 
Cam-clay plasticity Part III: Extension of the infinitesimal model to include finite strains. 
\CMAME 155, 73-95.

\bibitem{call}
Callari, C., Auricchio, F. and Sacco, E. (1998) 
A finite-strain Cam-clay model in the framework of multiplicative elasto-plasticity. 
\IJP 14, 1155-1187.

\bibitem{gajo}
Gajo, A., Bigoni, D. and Wood, D.M. (2004)
Multiple shear band development and related instabilities in granular materials. \JMPS in press.

\bibitem{Hill67}
Hill, R. (1967) 
On the classical constitutive laws for elastic/plastic solids.
In Broberg, B., ed., {\it Recent Progress in Applied Mechanics, The Folke Odkvist Volume}
Stockholm, Almqvist \& Wiksell. 241-249.

\bibitem{Hill68}
Hill, R. (1968) 
On constitutive inequalities for simple materials.
\JMPS 16, 229-242.

\bibitem{Hill78}
Hill, R. (1978) 
Aspects of invariance in solid mechanics. 
In Yih, C.-S., ed., {\it Advances in Applied Mechanics} 18. New York, Academic Press, 1-75.

\bibitem{HillRice}
Hill, R. and Rice, J. R. (1973) 
Elastic potentials and the structure of inelastic constitutive laws. 
{\it SIAM J.\ Appl.\ Math.} 25, 448-461.

\bibitem{hoger}
Hoger, A. (1987) 
The stress conjugate to logarithmic strain. 
\IJSS 23, 1645-1656.

\bibitem{ibr}
Ibrahimbegovic, A. (1994) Equivalent spatial and material descriptions of finite deformation 
elastoplasticity in principal axes. \IJSS 31, 3027-3040.

\bibitem{its}
Itskov, M. and Aksel, N. (2002) A closed-form representation for the derivative of non-symmetric tensor power 
series. \IJSS 39, 5963-5978.

\bibitem{lee}
Lee, E.H. (1969)
Elastic-plastic deformation at finite strains. 
\JAM 36, 1-6.

\bibitem{ogden84}
Ogden, R.W. (1984) 
{\it Non-linear elastic deformations}. Chichester, Ellis Horwood.

\bibitem{orti}
Ortiz, M. and Pandolfi, A. (2004) 
A variational Cam-clay theory of plasticity. 
\CMAME 193, 2645-2666.

\bibitem{peric}
Peric, D., Owen, D.R.J. and Honnor, M.E. (1992)
A model for finite strain elasto-plasticity based on logarithmic strains: Computational issues.
\CMAME 94, 35-61.

\bibitem{petryk}
Petryk, H. (2000)
General conditions for uniqueness in materials with multiple mechanisms of inelastic deformation. 
\JMPS 48, 367-396.

\bibitem{petr}
Petryk, H. and Thermann, K. (1985) 
Second-order bifurcation in elastic-plastic solids. \JMPS 33, 577-593.

\bibitem{picc}
Piccolroaz, A., Bigoni, D. and Gajo, A. (2005) An elastoplastic framework for granular materials
becoming cohesive through mechanical densification. Part I - small strain 
formulation of elastoplastic coupling.

\bibitem{rou}
Rouainia, M. and Muir Wood, D. (2000) 
An implicit constitutive algorithm for finite strain Cam-clay elasto-plastic model. 
{\it Mech. Cohes. Frict. Materials} 5, 469-489. 

\bibitem{sansour}
Sansour, C. (2001) 
On the dual variable of the logarithmic strain tensor, the dual variable of the Cauchy stress 
tensor, and related issues. 
\IJSS 38, 9221-9232.

\bibitem{sch}
Schieck, B. and Stumpf, H. (1993) Deformation analysis for finite elastic-plastic strains
in a Lagrangean-type description. \IJSS 30, 2639-2660.

\bibitem{simo}
Simo, J.C. and Meschke, G. (1993) 
A new class of algorithms for classical plasticity extended to finite strains. 
Application to geomaterials. 
{\it Comp. Mech.} 11, 253-278.

\bibitem{simo2}
Simo, J.C. and Miehe, C. (1992) Associative coupled thermoplasticity at finite strains: Formulation,
numerical analysis and implementation. \CMAME 98, 41-104. 

\bibitem{TruesdellNoll}
Truesdell, C. and Noll, W. (1965) The non-linear field theories
of mechanics. In Fl\"{u}gge, S., ed., {\it Encyclopedia of Physics}, III/3, 
Berlin, Springer-Verlag.

\bibitem{willis69}
Willis, J.R. (1969)
Some constitutive equations applicable to problems of large dynamic plastic deformation.
\JMPS 23, 405-419.

\end{thebibliography}
\end{document}